\documentclass[12pt]{article}
\usepackage{times}
\usepackage[dvips]{epsfig}

\newenvironment{sciabstract}{%
\begin{quote} \bf}
{\end{quote}}

\newcounter{lastnote}

\title{Entropically Driven Helix Formation \\ \normalsize {\sl final version appears in Science {\bf 307} (2005) 1067.} }

\author
{Yehuda Snir$^{1}$, Randall Kamien$^{1\ast}$\\
\\
\normalsize{$^{1}$Department of Physics and Astronomy, University of Pennsylvania,}\\
\normalsize{Philadelphia, PA, 19104, USA}\\
\\
\normalsize{$^\ast$To whom correspondence should be addressed; E-mail:  kamien@physics.upenn.edu.}
}

\date{12 October 2004; accepted 29 December 2004}

\begin{document} 

% Double-space the manuscript.

\baselineskip24pt

% Make the title.

\maketitle

% Place your abstract within the special {sciabstract} environment.

\begin{sciabstract}
The helix is a ubiquitous motif for biopolymers.  We propose a heuristic, entropically based
model which predicts helix formation in a system of hard spheres and semi-flexible tubes.  We find that the entropy of the spheres is maximized when short stretches of the tube form a helix with a
geometry close to that found in natural helices.  Our model could be directly tested with worm-like micelles as the tubes, and the effect could be used to self-assemble supramolecular helices.
\end{sciabstract}

% In setting up this template for *Science* papers, we've used both
% the \section* command and the \paragraph* command for topical
% divisions.  Which you use will of course depend on the type of paper
% you're writing.  Review Articles tend to have displayed headings, for
% which \section* is more appropriate; Research Articles, when they have
% formal topical divisions at all, tend to signal them with bold text
% that runs into the paragraph, for which \paragraph* is the right
% choice.  Either way, use the asterisk (*) modifier, as shown, to
% suppress numbering.

In the crowded
environment of the cell, long molecular chains frequently adopt
ordered, helical conformations.  Not only does this enable information
to be tightly packed, as in chromatin, but it also allows the machines
of transcription and repair to grapple on to a regular track.  Though geometrically
motivated models of folding ({\it 1\/}) with non-local interactions corroborate detailed models of proteins
({\it 2\/}), foldamers ({\it 3\/}), and DNA,  here we suggest a purely
entropic
approach to understand the folding of helices that exclusively relies on a local and
homogeneous interaction with depleting spheres.  This depletion interaction ({\it 4\/})
can be used as a surrogate for hydrophobicity, polymer-polymer interactions,
and for boundary layers in elastica and liquid crystals.  For a broad range of depletors, maximizing entropy leads to a universal polymer geometry; as the depletors grow large on the scale
of the polymer, this leads to
the favoring of sheet-like folding.

We model our polymer as a solid, impenetrable tube (radius $t$) immersed in a solution
of hard spheres (radius $r$) ({\it 5\/}).  The tube renders a region
of space inaccessible to the depleting spheres (Fig. 1A).  The spheres' entropy
increases as this excluded volume decreases by the overlap volume, $V_{\hbox{\small overlap}}$.  Though the size of the
excluded region is fixed by $r$ and $t$, the geometry can vary, allowing overlap of the inaccessible volumes from different parts
of the polymer,  increasing the entropy of the depleting spheres.  If the tube takes on a regular helix, the excluded volumes from adjacent turns and the central region overlap (Fig. 1B).  For small spheres
the tube will bend into a helix with a pitch (P) to radius (R) ratio $c^*=
2.5122$. This ratio is comparable to that seen, for instance, in many $\alpha$-helices of commonly
found proteins ({\it 1\/}) where $c\approx$ 5.4{\AA}/2.7{\AA}=2 is a lower bound.  The optimal ratio decreases as the colloid size grows, tending towards zero (Fig. 1C).   As $r/t$ grows large, the helix radius grows, causing a fixed length of tube to unwind.  This limit of infinite helix radius
is equivalent to a $\beta$-sheet since this is equivalent to
straight tubes lying in parallel.

As the colloid concentration increases the tendency to form
a helix grows strong enough to bend stiff polymers.  The persistence length
$\ell_P$ measures the relative stiffness of a rod and is the length scale over
which the rod can bend.  If a stiff rod of length $L$ bends into a helix with curvature $\kappa$, the total change in free energy, $\Delta F$, including the entropy of the colloids at concentration $n$ is
$\Delta F\propto {1\over 2} L\ell_P \kappa^2 - {n} V_{\hbox{\small overlap}}.$
Since the geometry of the helix relates $V_{\hbox{\small overlap}}/L$ to the curvature,
we can determine when $\Delta F<0$ and the depletion interaction dominates the
bending stiffness.  In Fig. 1D we plot the value of $\theta=nr^3/(\ell_P/t)$ at which the fully
collapsed helix is degenerate with a straight tube for a variety of colloid radii.   For segments of tube on the order of a few $\ell_P$ which cannot bend back on themselves, the helix should be the tightest packing, as with packing spheres in tubes ({\it 6\/}).

Our result suggests an explanation for the helical conformations at the onset of protein folding, the helical fiber-like growth of one liquid crystalline phase into another, and the helical supermolecular scaffold in condensed
chromatin, for instance.  It is straightforward to generalize this work to the induced interaction between
tubes.  In order to maximize the entropy of the colloidal depletants, the two tubes would form
helices of opposite handedness to achieve the tight packing known in liquid crystalline systems. Applied to the chromatin scaffold, the stereo-conformations of
sister chromatids in mitosis ({\it 7\/}) may be attributed to crowding in the cell.
As the density of polymers is increased,  $\ell_P/t$ controls their liquid crystalline ordering, while the volume fraction $nr^3$ of the colloids measures the intensity of the inter- and intra-tube attractions.  The interplay of these competing interactions, controlled by $\theta$, will lead to a rich phase diagram to explore through additional theory and experiment.

It has been observed that the helix content of polypeptides changes in
response to solvent ({\it 8\/}).  Though our model predicts trends and lends understanding, the details of complex interactions are encoded in effective values of  $r$ and $t$, much as effective masses and spring constants are used to fit the modes of crystals.  We note that our approach yields the same optimal geometry as equating the tube thickness
to its global radius of curvature ({\it 9\/}) and thus provides a novel, physical approach to study
the conformations of knotted and linked tubes.

{\bf References and Notes}
\begin{enumerate}
\item A. Maritan,C. Micheletti, A. Trovato, J.R. Banavar, {\sl Nature} {\bf 406}, 287
(2000).
\item K.F. Lau, K.A. Dill {\sl Macromolecules} {\bf 22}, 3986 (1989).
\item S.H. Gellman, {\sl Acc. Chem. Res.} {\bf 31}, 173 (1998).
\item S. Asakura, F. Oosawa, {\sl J. Chem Phys.} {\bf 22}, 1255 (1954).
\item J.R. Banavar, A. Maritan, {\sl Rev. Mod. Phys.} {\bf 75}, 23 (2003).
\item G.T. Pickett, M. Gross, H. Okuyama, {\sl Phys. Rev. Lett.} {\bf 85}, 3652 (2000).
\item E. Boy de la Tour, U.K. Laemmli, {\sl Cell} {\bf 55}, 937 (1988).
\item J. Bello, {\sl Biopolymers} {\bf 33}, 491 (1993).
\item O. Gonzalez, J.H. Maddocks, {\sl Proc. Natl. Acad. Sci. USA} {\bf 96}, 4769 (1999).
\item We thank G. Grason and D. Wu for discussions.  This work was supported by NSF Grants DMR01-02459, DMR01-29804 and the Pennsylvania Nanotechnology Institute.

\end{enumerate}

\clearpage

\begin{figure}\epsfxsize=6truein
\centerline{\epsfbox{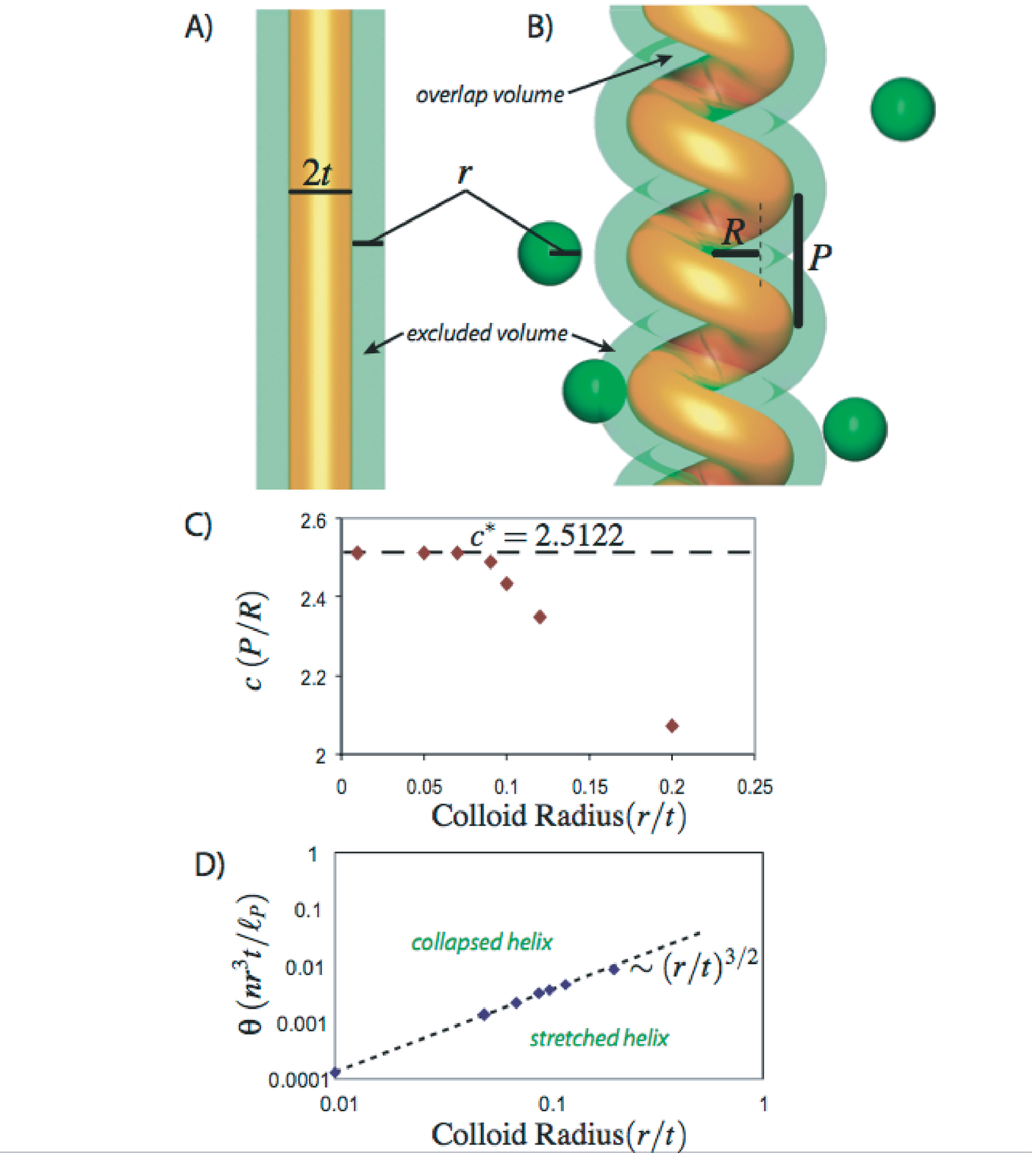}}

\caption{  A) Excluded volume (green) for a straight tube (orange).  
B) When the tube takes on a helical conformation the excluded volume self-overlaps,
reducing the total depleted volume.  C) The ratio $c=P/R$, which maximizes  $V_{\hbox{\small overlap}}$ vs. $r/t$.  The points were calculated by numerical integration of the overlap volume. D) The value of $\theta$ at which $\Delta F=0$ vs. $r/t$.  As $nr^3$ grows, the tube will go from a stretched helix to a collapsed (optimal) helix.  The points fall on the curve $(r/t)^{3/2}$, consistent
with the geometry (see Fig. S1).}
\end{figure}

\end{document}